# Phonon Coupled Scattering Caused Ultralow Lattice Thermal Conductivity and Its Role in The Remarkable Thermoelectric Performance of Newly Predicted SiS$_2$ and SiSe$_2$ monolayers

Jayanta Bera, Atanu Betal, Satyajit Sahu*

Department of Physics, Indian Institute of Technology Jodhpur, Jodhpur 342037, India

**Abstract:**

For high efficiency thermoelectric power conversion not only improvement of materials properties but also prediction and synthesis of new thermoelectric materials is needed. Here we have carried out a systematic investigation on thermoelectric performance of newly predicted two dimensional (2D) semiconducting SiS$_2$ and SiSe$_2$ monolayers of group IVA-VIA family using density functional theory (DFT) and Boltzmann transport equation (BTE). Our computed values of lattice thermal conductivity ($k_{ph}$) are ultralow which result very high thermoelectric figure of merit (ZT) value of 0.78 (0.80) at 500K in SiS$_2$ (SiSe$_2$) monolayer. The ultralow values of $k_{ph}$ are attributed to phonon-phonon coupling of acoustic and low frequency optical branches which leads to larger scattering, low group velocity, smaller mean free path and shorter lifetime of phonons. It is also found from our investigation that p-type doping is more effective than n-type doping to get optimal power factor (PF) and ZT. Our theoretical investigation suggests that newly predicted semiconducting SiS$_2$ and SiSe$_2$ monolayers can be very promising thermoelectric materials for fabrication of high efficiency thermoelectric power generator to convert wastage heat into electricity.

## 1. Introduction:

Generation of thermoelectric power is a new environment-friendly technique of converting wastage heat into electricity. But its lower efficiency of conversion of heat into electricity makes it lagged behind the conventional source of electric power such as thermal power, wind power, solar power etc[1]. So, search of materials with high thermoelectric PF and ZT to make optimal efficient thermoelectric power generator is a trending research now a days. The field of thermoelectric materials was mainly dominated by Skutterudites[2–6] Clathrates[7–10], complex alloys[11–13], metal chalcogenides and their alloys[14–19] and some oxides[20–22] in last few decades. Among these $Bi_2Te_3/Sb_2Te_3$ superlattice[23], $PbTe$[24], $SnSe$[25] and $Cu_2Se$[26] are well known thermoelectric materials for their high ZT value. After the discovery of 2D graphene people are interested in other two-dimensional layered materials such as transition metal dichalcogenides (TMDCs), group IVA-VIA and group IVA-dihalides compounds MXenes and black phosphorus owing to unique electronic, optical, mechanical and thermoelectric properties[27,28]. Two dimensional materials showed great potential in thermoelectric performance attributed to unique density of states (DOS) and fantastic combination of electrical conductivity (σ/τ) and $k_{ph}$ [29]. Theoretical and experimental investigation on well-known TMDCs $MoS_2$[30] and $WS_2$[31] monolayer suggest that in spite of high PF, the ZT is limited because of relatively high value of $k_{ph}$ as compared to other TMDCS such as $PtX_2$[32], $HfX_2$[33,34], $SnX_2$[35–37] and $ZrX_2$[38] (where x = S and Se), whereas ultralow values of $k_{ph}$ in $PtX_2$, $HfX_2$, $SnX_2$ and $ZrX_2$ result very high ZT greater than unity. Various efforts such as strain engineering[39–41] and chemical doping[42] have been attempted for the enhancement of thermoelectric efficiency. Although these two-dimensional layered materials showed good thermoelectric behavior but there are still some shortcomings to achieve high efficiency in thermoelectric power generation. So, the improvement in material properties as well as prediction and synthesis of new type of materials are needed to increase the thermoelectric power efficiency. Beyond the graphene and 2D TMDCs, group IVA-VIA family of 2D materials such as $SiS_2$, $SiSe_2$, $CS_2$ and $GeO_2$, made with light elements become very popular recently[43]. By using Ab initio calculation in association with molecular dynamics (MD) simulation electronic and structural properties of $SiS_2$ under pressure up to 100 GPa have been explored by Plašienka et al.[44]. Recent experimental synthesis[45] of layered $SiS_2$ with octahedral coordination at 7.5-9.0 GPa by laser heating of elemental Si and S at 1300K to 1700K revealed that these type of group IVA-VIA layered compounds can be synthesized. A very recent theoretical prediction[43] of $SiS_2$ and SiSe$_2$ monolayer and their electronic properties with applied strain has been studied which motivates us to investigate the thermoelectric performance of these two materials.

In this work, we have performed a systematic study of structural, electronic properties and for the first time ever known we have investigated thermoelectric performance and lattice thermal conductivity ($k_{ph}$) in SiS$_2$ and SiSe$_2$ monolayer with the help of DFT and BTE. Our calculated values of $k_{ph}$ in SiS$_2$ and SiSe$_2$ are 5.43 W/(mK) and 1.675 W/(mK) at 300K respectively which result a large ZT value of 0.66 (0.73) at 300K and 0.78 (0.80) at 500K in SiS$_2$ (SiSe$_2$) monolayer. It has also been observed that p-type doping is favorable for optimal thermoelectric performance in these two materials. The ultralow values of $k_{ph}$ are attributed to phonon-phonon coupling which leads to larger scattering, low group velocity, smaller mean free path and shorter lifetime of phonons. Our theoretical calculation of high ZT and ultralow $k_{ph}$ values suggests that newly predicted SiS$_2$ and SiSe$_2$ monolayers can have great potential to fabricate high efficiency thermoelectric power generator to convert wastage heat into electricity.

## 2. Computational Details:

First principle calculations were carried out using DFT [46] in Quantum Espresso (QE)[47] package. Ultrasoft pseudopotential for the electron-ion interactions and Perdew-Burke-Ernzerhof (PBE)[48] generalized gradient approximation (GGA)[49] as electronic exchange and correlation functional have been used throughout the calculations. A 18×18×1 dense k point sampling in hexagonal Brillouin zone (BZ) were used for the geometry relaxation of the unit cell whereas the cutoff energy for the plane wave function was kept at 50 Ry. For non-self-consistent (nscf) calculations a 60×60×1 dense k point sampling in BZ were used. Geometry relaxations were performed until the maximum force in individual atom reduced to a value of 0.01eV/Å. The effect of monolayer was implemented by creating sufficient vacuum of length 17 Å along the c direction to avoid interlayer interactions.

The thermoelectric properties have been calculated using BTE as implemented in BoltzTrap[50] code considering constant scattering time approximation (CSTA), which enables temperature and doping dependent electronic and thermal transport properties calculations assuming scattering time is unchanged with the energy. The details of the thermoelectric properties calculation can be found in our previous work[39]. The linearized phonon Boltzmann equation was used to calculate the $k_{ph}$ in Phono3py[51] interfaced to QE package[47]. To calculate 3$^{rd}$ order anharmonic and 2$^{nd}$ order harmonic interatomic force constant 2×2×1 supercell was used with 9×9×1 k point sampling and for calculating $k_{ph}$ 96×96×1 k point sampling was used.

## 3. Results and discussions:

### 3.1. Structural parameters:

The unit cell of SiS$_2$ and SiSe$_2$ monolayers have a 1T-CdI$_2$ type crystal structures belonging to space group $P\bar{3}m1(164)$ where S or Se atoms are buckled in two different planes whereas Si atoms are sandwiched between two S or Se planes as shown in fig. 1a and fig. 1b. The optimized lattice constants of 3.30 Å (SiS$_2$) and 3.51 Å (SiSe$_2$) have been obtained from DFT with GGA-PBE functional. The Si-S bond length and S-Si-S bond angle have been calculated to be 2.312 Å and 89.5º in SiS$_2$ whereas Si-Se bond length and Se-Si-Se bond angle have been found to be 2.485 Å and 90.15º in SiSe$_2$ monolayer. Our calculated values of lattice constant, bond lengths and bond angles match very well with previous computed data[43].

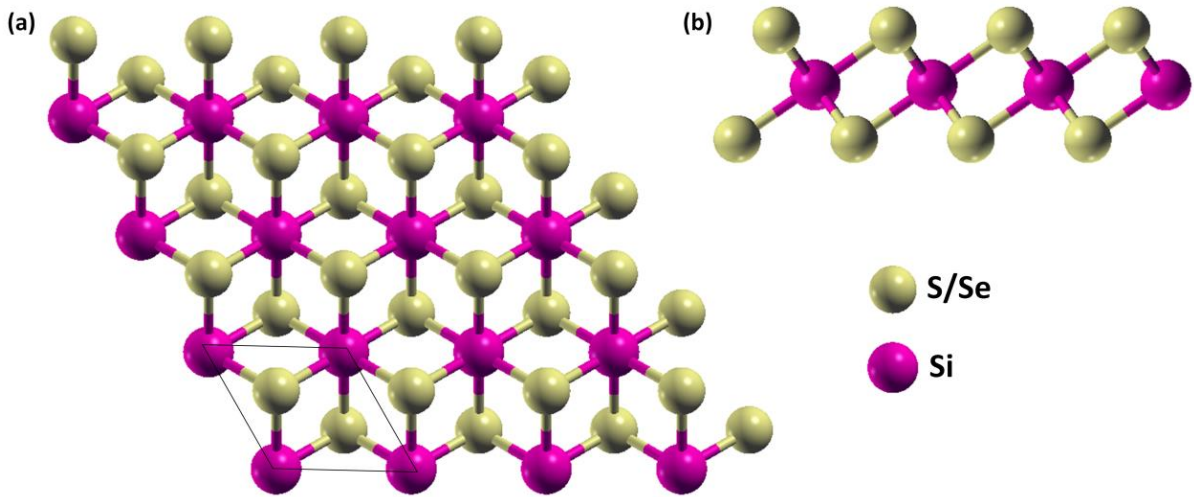

**Fig. 1.** (a) Top view and (b) side view of a 4×4×1 supercell SiX$_2$ (X=S/Se) monolayer. The hexagonal unit cell is shown by black box.

### 3.2. Structural stability:

*3.2.1. Cohesive energy:*

To find the structural stability of our structures we have calculated the cohesive energy per atom ($E_C$) using the expression

$$E_C = [E_{Si} + 2E_{S/Se} - E_{SiS_2/SiSe_2}]/3, \qquad (1)$$

where $E_{Si}$ and $E_{S/Se}$ are the total ground state energy of a single Si atom and a single S or Se atom and $E_{SiS_2/SiSe_2}$ is the total ground state energy of the SiS$_2$ or SiSe$_2$ unit cell. From our calculations cohesive energy per atom of SiS$_2$ and SiSe$_2$ has been found with a value of 5.08 eV/atom and 4.46 eV/atom respectively which is in good agreement with previous calculation[43]. Our calculated values are comparable with other 2D materials such as graphene (8.00 eV/atom)[52], silicene (3.94 eV/atom)[53], phosphorene (3.44 eV/atom)[54], WS$_2$ (2.87 eV/atom)[39] and Be$_2$C monolayer (4.84 eV/atom)[55] which suggest very good structural stabilities of the considered structures.

*3.2.2. Phonon dispersion curve:*

Phonon dispersion curves of SiS$_2$ and SiSe$_2$ have been plotted along the hexagonal BZ path Γ-M-K-Γ as shown in Fig. 2a and Fig. 2b respectively. There is a very small frequency domain in the negative side (-2 cm$^{-1}$ in SiS$_2$ and -3 cm$^{-1}$ in SiSe$_2$ at Γ point) which suggests that the structures are dynamically stable. The highest frequencies of 17.24 THz (575 cm$^{-1}$) and 14 THz (467 cm$^{-1}$) have been calculated for SiS$_2$ and SiSe$_2$ respectively and it is clear that SiSe$_2$ has lower frequency than that of SiS$_2$. It is also seen that the acoustic branches are coupled to the nearest optical branches in both materials. The acoustic branches are contributed mainly by the S or Se atoms whereas optical branches are contributed by combination of Si and S or Se atoms as seen from the projected PhDOS plot. Since, Se atoms are heavier than S atoms and will vibrate with lower frequency than that of S atoms hence SiSe$_2$ monolayer has lower phonon frequency than that of SiS$_2$ monolayer.

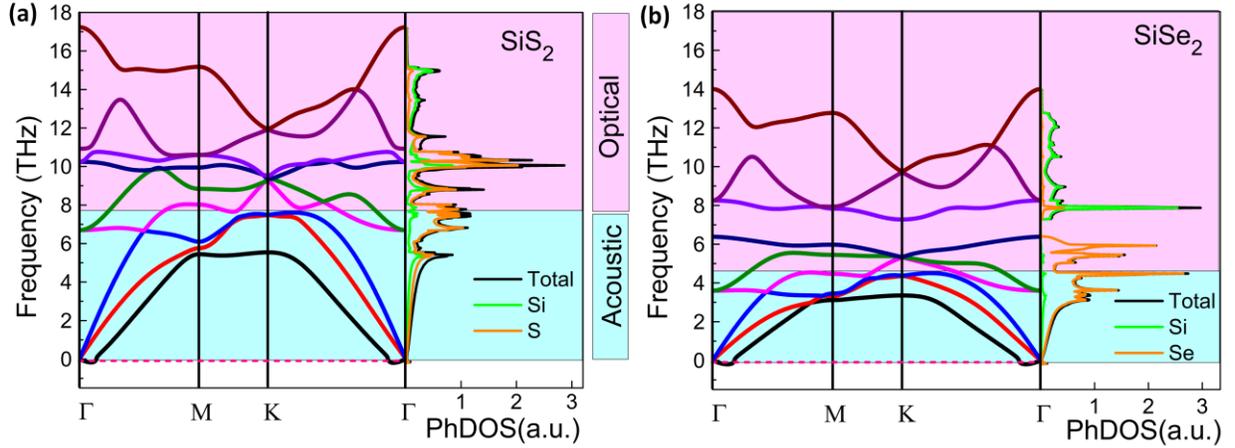

**Fig. 2.** Phonon dispersion curves and PhDOS of (a) SiS$_2$ and (b) SiSe$_2$. The black, red and blue line in the dispersion curves act for out of plane acoustic (ZA), transverse acoustic (TA) and longitudinal acoustic (LA) modes respectively.

*3.3. Electronic band structures:*

The electronic band structures of monolayer SiS$_2$ and SiSe$_2$ have been calculated using GGA-PBE along high symmetry path Γ-M-K-Γ in hexagonal BZ as shown in Fig. 3a and Fig. 3b. Both the structures show semiconducting indirect band gap as conduction band minima (CBM) lies on M point whereas valance band maxima (VBM) lies in between high symmetry points Γ and M. An indirect band gap of 1.39 eV and 0.48 eV have been calculated in SiS$_2$ and SiSe$_2$ monolayer respectively which are very close to the values with previous PBE calculations but lower as compared to the band gap calculated[43] (2.45 eV for SiS$_2$ and 1.37 eV for SiSe$_2$) with HSE06 functional. This is because PBE underestimates the band gap. To find the contribution of Si and S or Se atom on the electronic properties of SiS$_2$ and SiSe$_2$ monolayers, projected DOS (PDOS) has been calculated as shown in Fig. 3c and Fig. 3d. As seen from total DOS of both the structures it is clear that density of states is higher in valance band as compared to conduction band. The gap between valance band states and conduction band states are much higher in SiS$_2$ as compared to SiSe$_2$ as band gap is higher in SiS$_2$. It is also clear that both valance and conduction band are mostly dominated by 2p and 3p orbitals of chalcogenide S and Se atoms respectively whereas very small contribution of Si atoms near the band edges has been observed in SiS$_2$ and SiSe$_2$ monolayers. We have also plotted two-dimensional charge density as shown in Fig. 4 where one can clearly see that charges are localized mostly on S or Se atoms and less amount of charge on Si atoms. Therefore, S (Se) plays significant role than that of Si atoms in electronic properties of $SiS_2$ ($SiSe_2$) monolayers. It is also noted, that strong polar covalent type of Si-S(Se) bonding in these monolayers is owing to the electronegativity difference of Si, S and Se atoms.

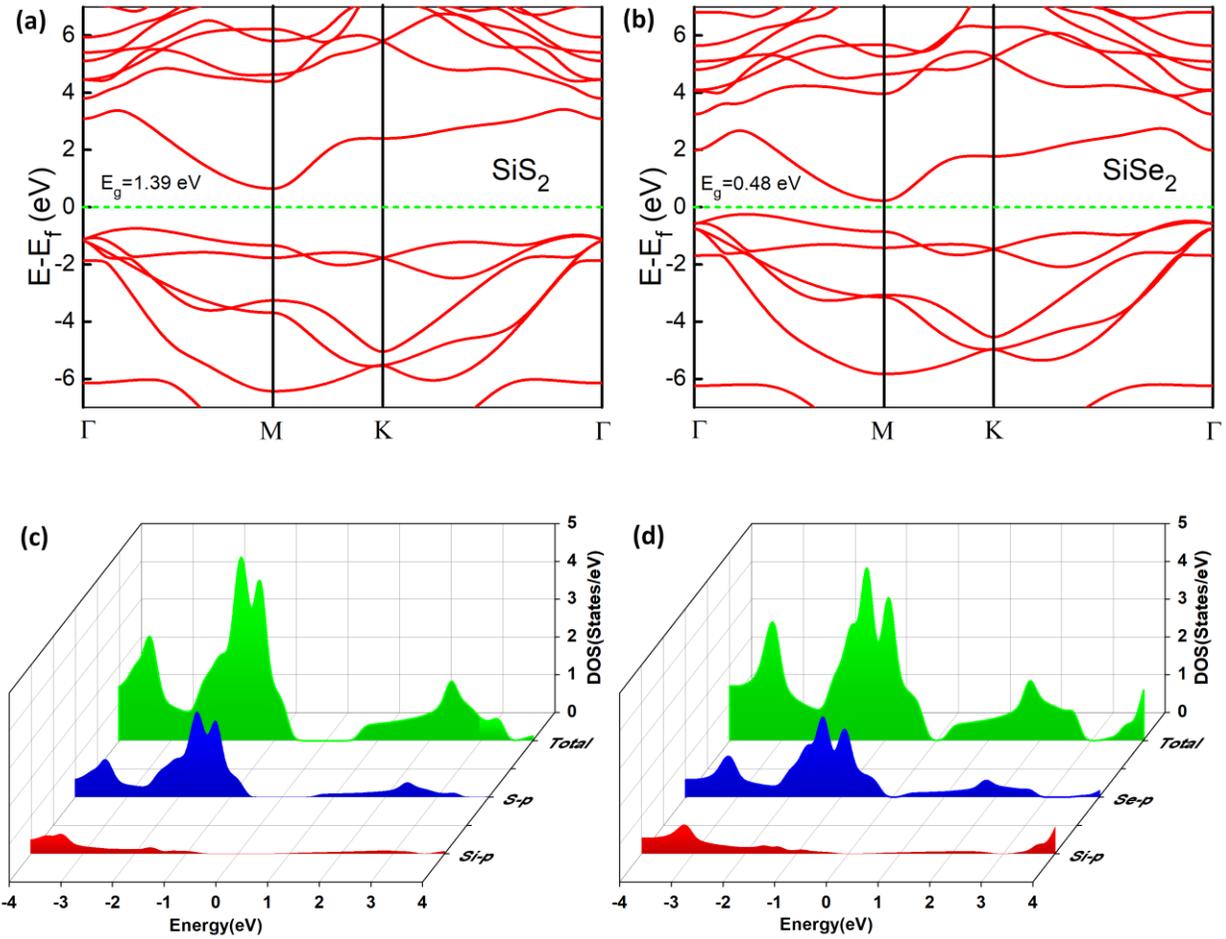

**Fig. 3.** Electronic band structures of (a) $SiS_2$ (b) $SiSe_2$ monolayers with PBE. PDOS of (c) $SiS_2$ (d) $SiSe_2$ monolayers.

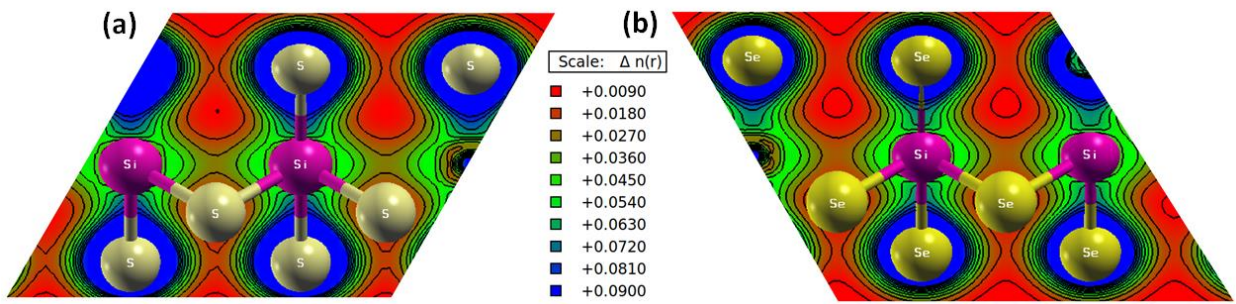

**Fig. 4.** Two-dimensional charge density plot of (a) SiS$_2$ and (b) SiSe$_2$ monolayers.

## *3.4. Carrier's mobility and relaxation time:*

To calculate the mobility of carriers in the structures, deformation potential ($E_l$) theory as proposed by Bardeen and Shockley[56] has been used. As stated in this theory, mobility of two-dimensional materials ($\mu_{2D}$) can be determined as

$$\mu_{2D} = \frac{e\hbar^3 C_{2D}}{K_B T m^* m_d^* (E_l^2)} \qquad (2)$$

where $\hbar$, $K_B$, e and $T$ are the Planck's constant, Boltzmann constant, elementary charge and temperature respectively. The elastic constant $C_{2D}$ is defined as $C_{2D} = [\partial^2 E/\partial \varepsilon^2]/A_0$ where E, $\varepsilon$ and $A_0$ are total energy, amount of uni-axial strain and the optimized cell's area respectively. $m^* = \hbar^2/[\partial^2 E/\partial k^2]$ is the effective mass along the propagation direction and $m_d^* = \sqrt{m_\parallel m_\perp}$ where $m_\parallel (m_\perp)$ is the effective mass parallel (perpendicular) to the propagation direction. $E_l$ can be obtained for electron and hole by calculating the amount of shift in CBM and VBM in terms of energy with applied strain as shown in Fig. 5c and 5d. The relaxation time can be determined in terms of mobility as $\tau = m^* \mu/e$. The effective mass of the electron and hole are anisotropic in nature in SiS$_2$ and SiSe$_2$. Our theoretical calculation reveals that SiS$_2$ and SiSe$_2$ have high carrier mobility of the order of ~$10^3$ cm$^2$V$^{-1}$s$^{-1}$. Our calculated values of mobility are higher than mobility of popular two-dimensional TMDC MoS$_2$ (200 cm$^2$V$^{-1}$s$^{-1}$)[57] and comparable to other 2D materials such as phosphorene (1.0×10$^3$ cm$^2$V$^{-1}$s$^{-1}$)[58], GeS monolayer (3.7×10$^3$ cm$^2$V$^{-1}$s$^{-1}$)[59] and GeTe monolayer (1.0×10$^3$ cm$^2$V$^{-1}$s$^{-1}$)[60]. All the parameters related to mobility are listed in Table 2.

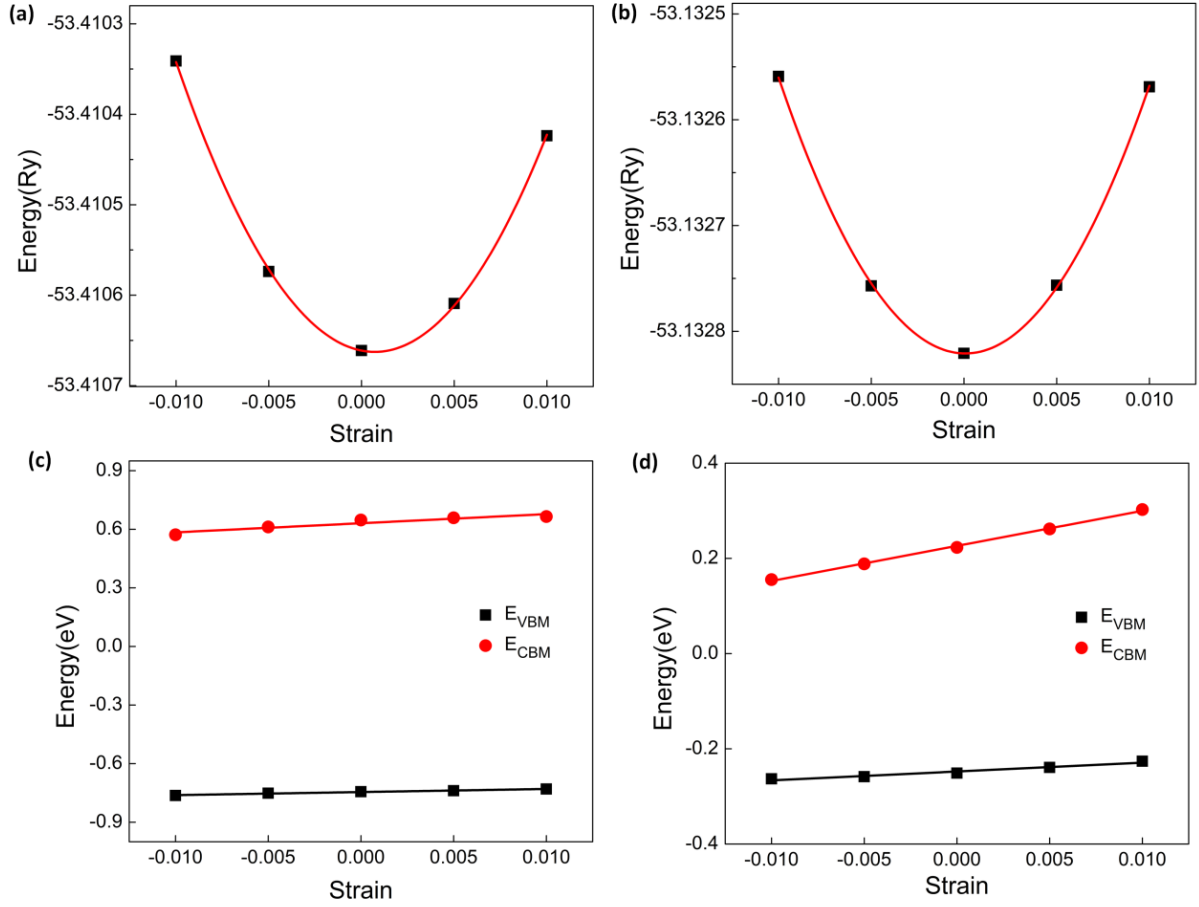

**Fig. 5.** Variation of total energy with uni-axial strain in (a) SiS$_2$ and (b) SiSe$_2$. Shift in the CBM and VBM in eV with uni-axial strain in (c) SiS$_2$ and (d) SiSe$_2$ to determine deformation potential.

| Sample | Carrier | $m^*_{\Gamma-M}$ | $m^*_{K-M}$ | $C_{2D}$ (N/m) | $E_l$ (eV) | $\mu_{2D}$ (cm$^2$V$^{-1}$s$^{-1}$) | $\tau$ (s)×10$^{-13}$ |
|---|---|---|---|---|---|---|---|
| SiS$_2$ | electron | 0.43 | 0.14 | 129 | 4.678 | 3664 | 3.00 |
|  | hole | 0.54 | 1.25 | 129 | 1.577 | 1079 | 7.68 |
| SiSe$_2$ | electron | 0.41 | 0.14 | 105 | 7.359 | 1233 | 0.98 |
|  | hole | 0.52 | 0.93 | 105 | 1.864 | 997 | 5.30 |

**Table 2.** Calculated values of effective mass, $C_{2D}$, $E_l$, $\mu_{2D}$ and $\tau$ of electrons and holes in monolayer SiS$_2$, SiSe$_2$.

### 3.5. Thermoelectric performance:

To find the thermoelectric performance of monolayer $SiS_2$ and $SiSe_2$, the variation of Seebeck coefficient (S), relaxation time scaled electrical conductivity (σ/τ), electronic thermal conductivity ($K_{el}/\tau$) and PF($S^2\sigma/\tau$) as a function of chemical potential (μ) at 300K to 500K have been plotted in Fig. 6a and Fig. 6b respectively. The highest S value of 2100 µV/K (n-type) and 2281 µV/K (p-type) in $SiS_2$ and 654 µV/K (n-type) and 802 µV/K (p-type) in $SiSe_2$ at 300K has been calculated. S is higher in $SiS_2$ than that of $SiSe_2$ because S depends on band gap and $SiSe_2$ has a lower band gap. As temperatures rises the value of S drops whereas σ/τ and $K_{el}/\tau$ increase with rising temperature. The thermoelectric power factor (PF) has been calculated to be 15.61 $Wm^{-1}K^{-2}s^{-1}$ (p-type) and 7.71 $Wm^{-1}K^{-2}s^{-1}$ (n-type) in $SiS_2$ and 15.55 $Wm^{-1}K^{-2}s^{-1}$ (p-type) and 7.77 $Wm^{-1}K^{-2}s^{-1}$ (n-type) in $SiSe_2$ monolayer at 300K. So, it is obvious that the PF for p-type carriers are almost double that of n-type carriers in both the materials at 300K suggesting p-type semiconducting nature of these monolayers. The reason of this type of behavior can be found from the total DOS as shown in Fig. 3c and Fig. 3d where it is clearly seen that the number of energy states per eV in the valance band are much higher than that in conduction band. At 500K, highest PF of 22.96 $Wm^{-1}K^{-2}s^{-1}$ and 23.04 $Wm^{-1}K^{-2}s^{-1}$ for p-type carriers in $SiS_2$ and $SiSe_2$ monolayers respectively has been obtained. So, for both materials S and PF for p-type is much higher than that of n-type which suggest p-type doping is more effective than n-type to get optimum thermoelectric performance in these monolayers. In contrast, n-type doping is preferable for thermoelectric application in popular two-dimensional TMDCs thermometric materials such as $MoS_2$[61] and $WS_2$[39].

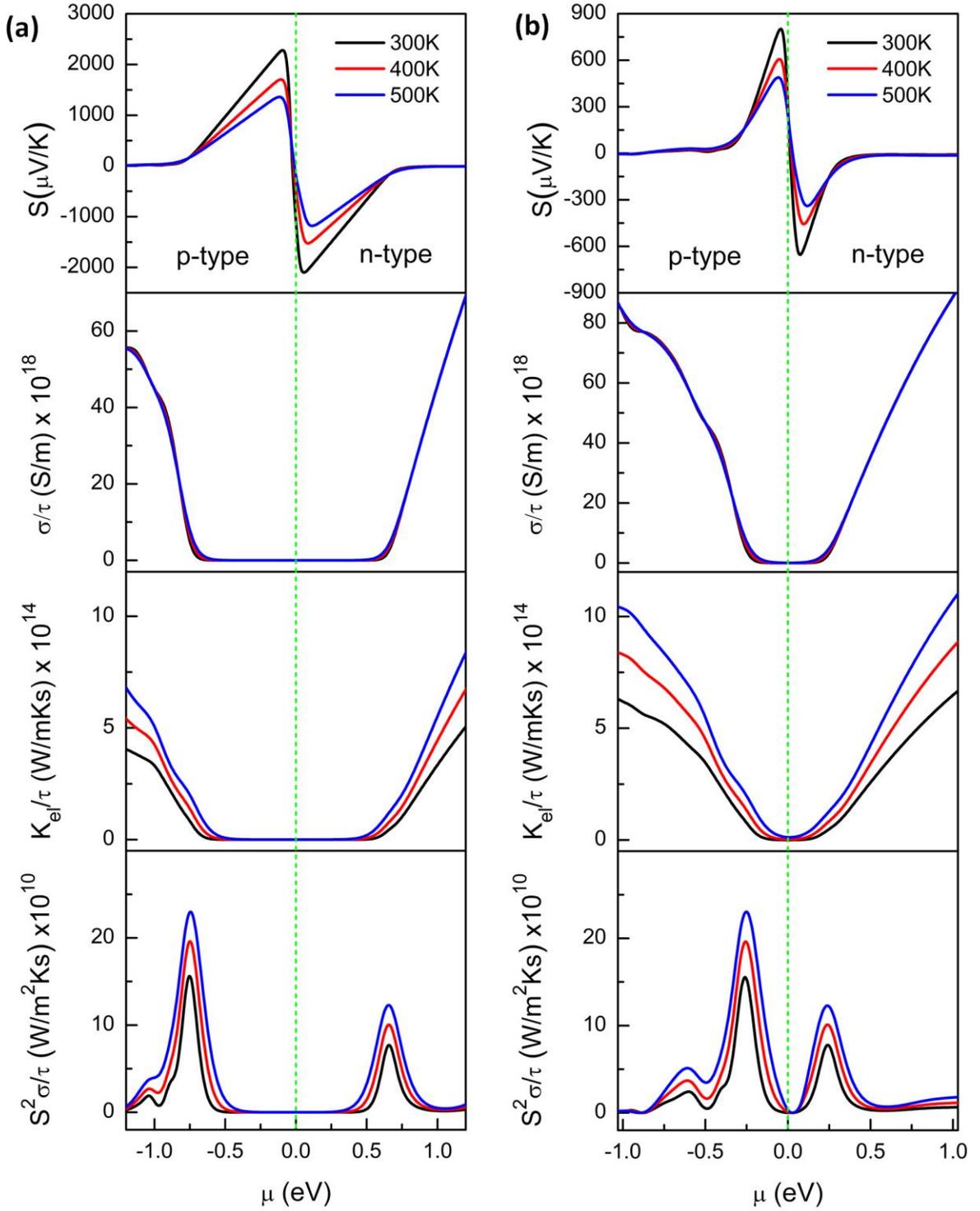

**Fig. 6.** Variation of S, σ/τ, $K_{el}/\tau$ and $S^2\sigma/\tau$ as a function of μ at 300K to 500K in (a) $SiS_2$ and (b) $SiSe_2$ monolayers. Fermi level is shown by green dashed lines.

### 3.6. Lattice thermal conductivity ($k_{ph}$):

The variation of $k_{ph}$ with temperature in SiS$_2$ and SiSe$_2$ has been plotted in Fig. 7a and $k_{ph}$ varies as T$^{-1}$ with temperature owing to phonon-phonon scattering. Our calculated values of $k_{ph}$ in SiS$_2$ and SiSe$_2$ are 5.43 W/(mK) and 1.675 W/(mK) at 300K respectively. Our calculated ultralow values of $k_{ph}$ are comparable with well-known high-performance thermoelectric materials such as Bi$_2$Te$_3$ ($k_{ph}$=1.6 W/mK)[62], PbTe ($k_{ph}$=2.2 W/mK)[63] and SnSe ($k_{ph}$=2.6 W/mK)[64] and much lower than popular 2D TMDCs MoS$_2$ ($k_{ph}$=34.5 W/mK)[65] and WS$_2$ ($k_{ph}$=72 W/mK)[39]. To investigate the origin of ultralow values of $k_{ph}$ we have to look into the phonon dispersion curves as shown previously in Fig. 2. It is clear from phonon dispersion curves of both the materials that unlike MoS$_2$ and WS$_2$, there is no gap between acoustic and optical branches and the acoustic branches are coupled to the optical branches in monolayer SiS$_2$ and SiSe$_2$. In MoS$_2$, WS$_2$, MoSe$_2$ and WSe$_2$ [66,67] there is a finite gap between acoustic and optical branches. Due to this coupling phonon-phonon scattering increases and group velocity (v$_g$) and mean free path (MFP) of phonons decrease which induce low values of $k_{ph}$ in SiS$_2$ and SiSe$_2$ monolayer.

The cumulative $k_{ph}$ and its derivative have been plotted as a function of phonon frequency and shown in Fig. 7b. It is clearly seen that $k_{ph}$ becomes almost constant around 6 THz in SiSe$_2$ whereas it became constant around 10 THz in SiS$_2$. So, in both the monolayers acoustic modes which are in the lower frequency region (as shown by cyan colour in Fig. 2) have more effect on $k_{ph}$ than that of optical modes. Also, phonons have lower frequency in SiSe$_2$ than in SiS$_2$ because of larger atomic mass of Se than S atoms. To find the contribution of phonons with different MFP we have plotted cumulative $k_{ph}$ divided by total $k_{ph}$ as a function of phonon MFP as shown in Fig. 7c. where the blue and red colours have been used for SiS$_2$ and SiSe$_2$ respectively. It has been observed that almost 90% of the total $k_{ph}$ values are contributed by those phonons which have MFP less than 623Å in SiS$_2$ and 185Å in SiSe$_2$ monolayer. Therefore, it is clear that phonons have much smaller MFP and which will result stronger scattering and induce lower $k_{ph}$ in SiSe$_2$ than that in SiS$_2$.

The mode dependent $v_g$ as a function of phonon frequency has been plotted in Fig. 7d and Fig. 7e for SiS$_2$ and SiSe$_2$ monolayer respectively. The group velocity of both acoustic and optical phonon branches is lower in SiSe$_2$ than that of SiS$_2$ which means the phonons move with smaller group velocity in SiSe$_2$. The smaller group velocity of phonons also induces lower $k_{ph}$ in SiSe$_2$ monolayers. Almost 90% of the $k_{ph}$ is contributed by acoustic branches and 10% by optical branches in both the monolayers as shown in Fig. 8. ZA branches contribute more among the acoustic branches owing to the lower group velocity of ZA branch than TA and LA branch. Phonon lifetime as a function of frequency has been shown in Fig. 9. and it is observed that phonon lifetime of SiSe$_2$ is shorter than SiS$_2$. Therefore, lower group velocity, smaller MFP and shorter lifetime of phonons result stronger scattering of phonons which induce lower value of $k_{ph}$ in SiSe$_2$ than that of SiS$_2$.

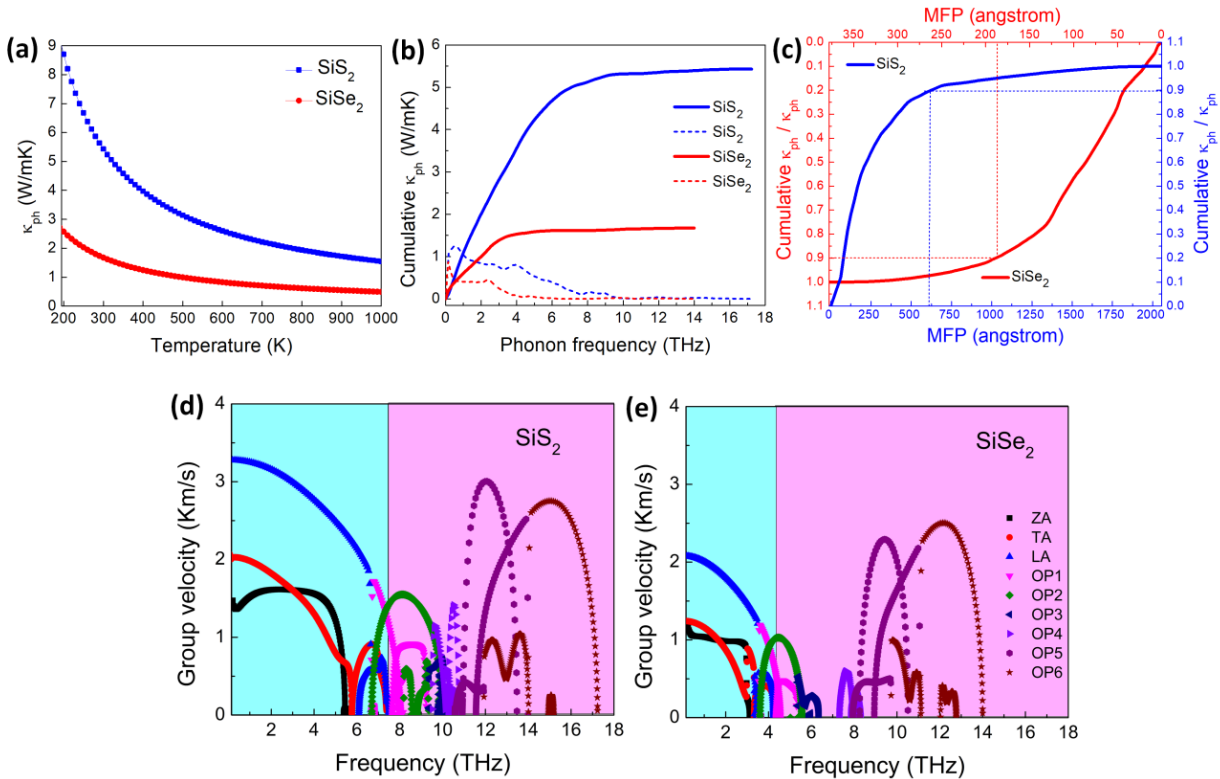

**Fig.7.** (a) Variation of $k_{ph}$ with temperature (b) cumulative $k_{ph}$ and its derivative (shown by dashed lines) as a function of phonon frequency and (c) cumulative $k_{ph}$ divided by total $k_{ph}$ as a function of MFP in SiS$_2$ and SiSe$_2$ monolayer. Group velocity of different phonon modes as a function of frequency in (d) SiS$_2$ and (e) SiSe$_2$ where cyan and magenta colored regions represent acoustic and optical branches respectively.

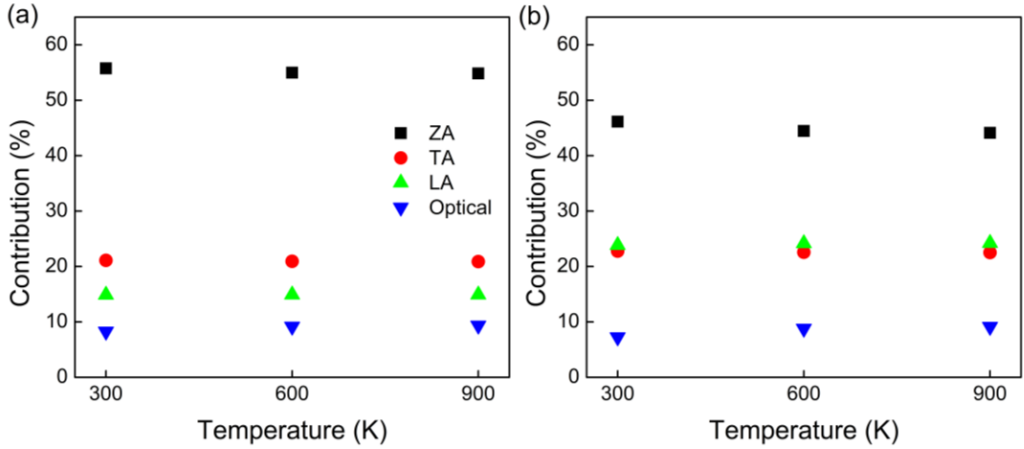

**Fig.8.** Percentage contribution of different phonon modes to the total $k_{ph}$ in (a) SiS$_2$ and (b) SiSe$_2$.

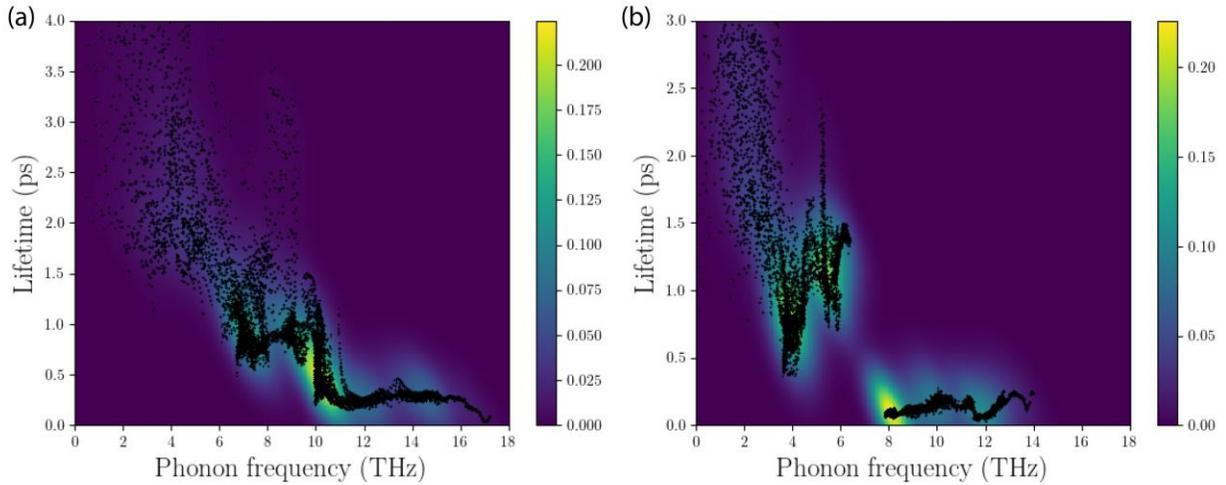

**Fig. 9.** Phonon lifetime in pico-second as a function of phonon frequency in Tera-Hertz in (a) SiS$_2$ and (b) SiSe$_2$ monolayers.

*3.7. Thermoelectric figure of merit (ZT):*

The term ZT defines the quality of a thermoelectric material and can be expressed by

$$ZT = \frac{S^2 \sigma T}{K_{el} + k_{ph}}$$

where all the parameters are defined in previous sections. The variation of ZT as a function of μ (eV) in SiS$_2$ and SiSe$_2$ are shown in Fig. 10a and Fig. 10b respectively. We have calculated ZT value of 0.66 in SiS$_2$ and 0.73 in SiSe$_2$ respectively at 300K for p-type carriers. These values are very high at 300K and we have calculated highest ZT value of 0.78 and 0.80 in SiS$_2$ and SiSe$_2$ respectively for p-type carriers at 500K. All the ZT values are listed in Table 3. Like PF values, ZT values are also higher for p-type carriers than n-type, which indicates that p-type doping is favorable for optimal thermoelectric performance in SiS$_2$ and SiSe$_2$ monolayers to achieve maximum ZT. SiSe$_2$ has slightly higher ZT value as compared to SiS$_2$, because of lower value of $k_{ph}$ owing to lower group velocity, lower MFP and stronger scattering in SiSe$_2$ as discussed in previous section. So, these high values of ZT at room temperature (300K) and slightly higher temperature (500K) reveals that monolayer SiS$_2$ and SiSe$_2$ can be used as highly efficient thermoelectric materials operating near room temperature.

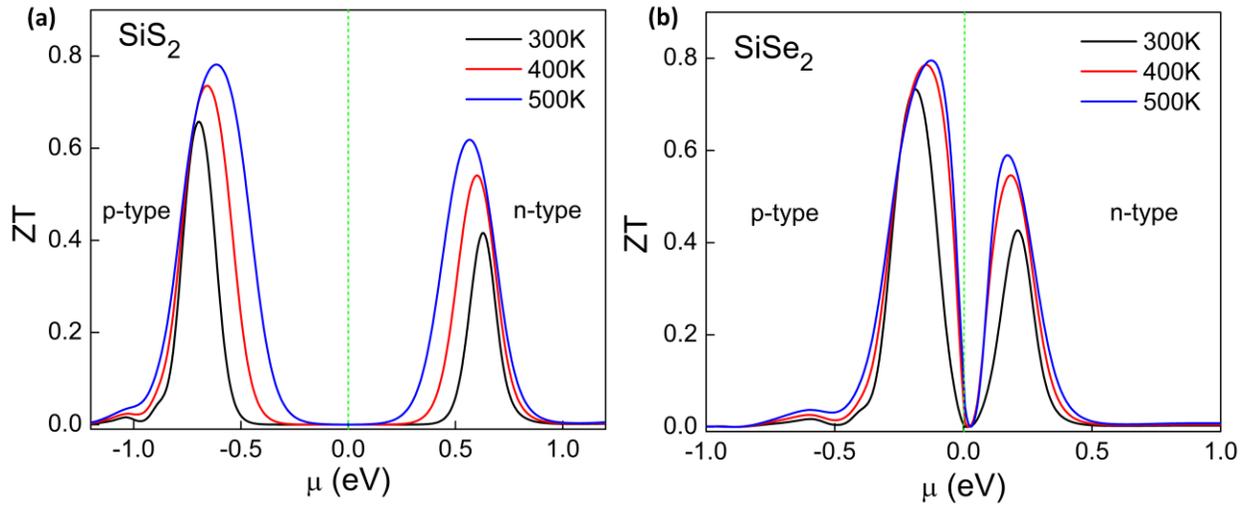

**Fig. 10.** Variation of ZT as a function of μ at 300K to 500K in (a) SiS$_2$ and (b) SiSe$_2$ monolayers.

| Sample | | S (μV/K) | | | S$^2$σ/τ (W/mK$^2$s) ×10$^{10}$ | | | ZT | | |
|---|---|---|---|---|---|---|---|---|---|---|
| | | 300K | 400K | 500K | 300K | 400K | 500K | 300K | 400K | 500K |
| SiS$_2$ | n | 2100 | 1524 | 1180 | 7.71 | 10.03 | 12.29 | 0.42 | 0.54 | 0.62 |

|  | p | 2281 | 1706 | 1361 | 15.61 | 19.60 | 22.96 | 0.66 | 0.73 | 0.78 |
| --- | --- | --- | --- | --- | --- | --- | --- | --- | --- | --- |
| SiSe$_2$ | n | 654 | 457 | 339 | 7.77 | 10.10 | 12.28 | 0.43 | 0.55 | 0.59 |
|  | p | 802 | 606 | 488 | 15.55 | 19.61 | 23.04 | 0.73 | 0.78 | 0.80 |

**Table 3.** Calculated values of S, S$^2\sigma/\tau$ and ZT at 300K, 400K and 500K in monolayer SiS$_2$ and SiSe$_2$.

## 4. Conclusions:

In summary, a systematic investigation of electronic, structural and thermoelectric properties of newly predicted monolayers SiS$_2$ and SiSe$_2$ has been carried out with the help of DFT and BTE. Our calculated ultralow values of $k_{ph}$ at 300K in SiS$_2$ and SiSe$_2$ are 5.43 W/(mK) and 1.675 W/(mK) respectively. These ultralow values of $k_{ph}$ are attributed to phonon-phonon coupling of acoustic and optical branches which leads to larger scattering, low group velocity, smaller mean free path and shorter lifetime of phonons. High PF for p-type carriers and ultralow $k_{ph}$ result high ZT values of 0.78 (0.80) at 500K in SiS$_2$ (SiSe$_2$) monolayer. From our investigation it has been found that p-type doping is more effective than n-type doping in both the materials of our interest to get optimal ZT and PF. Our calculated carrier mobility values are high which are in the orders of ~10$^3$ cm$^2$V$^{-1}$s$^{-1}$ for these two materials. From our investigation it is clear that newly predicted semiconducting monolayer SiS$_2$ and SiSe$_2$ could be very promising new generation thermoelectric materials for fabrication of high efficiency thermoelectric power generator to convert wastage heat into electricity.

## Conflicts of interest

There are no conflicts to declare.

## Acknowledgement:

Cu2Se1-xIx, *Adv. Mater.*, 2013, **25**, 6607–6612.

27  S. Z. Butler, S. M. Hollen, L. Cao, Y. Cui, J. A. Gupta, H. R. Gutiérrez, T. F. Heinz, S. S. Hong, J. Huang and A. F. Ismach, Progress, challenges, and opportunities in two-dimensional materials beyond graphene, *ACS Nano*, 2013, **7**, 2898–2926.

28  A. Betal, J. Bera and S. Sahu, Low-temperature thermoelectric behavior and impressive optoelectronic properties of two-dimensional XI2 (X= Sn, Si): A first principle study, *Comput. Mater. Sci.*, 2021, **186**, 109977.

29  D. Li, Y. Gong, Y. Chen, J. Lin, Q. Khan, Y. Zhang, Y. Li, H. Zhang and H. Xie, Recent Progress of Two-Dimensional Thermoelectric Materials, *Nano-Micro Lett.*, 2020, **12**, 36.

30  K. Hippalgaonkar, Y. Wang, Y. Ye, D. Y. Qiu, H. Zhu, Y. Wang, J. Moore, S. G. Louie and X. Zhang, High thermoelectric power factor in two-dimensional crystals of Mo S 2, *Phys. Rev. B*, 2017, **95**, 115407.

31  W. Huang, H. Da and G. Liang, Thermoelectric performance of MX2 (M Mo,W; X S,Se) monolayers, *J. Appl. Phys.,* 2013, ***113***, 104304.

32  J. Zhang, Y. Xie, Y. Hu and H. Shao, Remarkable intrinsic ZT in the 2D PtX2 (X= O, S, Se, Te) monolayers at room temperature, *Appl. Surf. Sci.*, 2020, **532**, 147387.

33  M. K. Mohanta, A. Rawat, N. Jena, R. Ahammed and A. De Sarkar, Ultra-low lattice thermal conductivity and giant phonon–electric field coupling in hafnium dichalcogenide monolayers, *J. Phys. Condens. Matter*, 2020, **32**, 315301.

34  F. Khan, H. U. Din, S. A. Khan, G. Rehman, M. Bilal, C. V Nguyen, I. Ahmad, L.-Y. Gan and B. Amin, Theoretical investigation of electronic structure and thermoelectric properties of MX2 (M= Zr, Hf; X= S, Se) van der Waals heterostructures, *J. Phys. Chem. Solids*, 2019, **126**, 304–309.

35  G. Li, G. Ding and G. Gao, Thermoelectric properties of SnSe2 monolayer, *J. Phys. Condens. Matter*, 2016, **29**, 15001.

36  J. Li, J. Shen, Z. Ma and K. Wu, Thickness-controlled electronic structure and
20

transistors, *Nat. Nanotechnol.*, 2011, **6**, 147–150.

58  S. Das, W. Zhang, M. Demarteau, A. Hoffmann, M. Dubey and A. Roelofs, Tunable transport gap in phosphorene, *Nano Lett.*, 2014, **14**, 5733–5739.

59  F. Li, X. Liu, Y. Wang and Y. Li, Germanium monosulfide monolayer: a novel two-dimensional semiconductor with a high carrier mobility, *J. Mater. Chem. C*, 2016, **4**, 2155–2159.

60  M. Qiao, Y. Chen, Y. Wang and Y. Li, The germanium telluride monolayer: a two dimensional semiconductor with high carrier mobility for photocatalytic water splitting, *J. Mater. Chem. A*, 2018, **6**, 4119–4125.

61  Dimple, N. Jena and A. De Sarkar, Compressive strain induced enhancement in thermoelectric-power-factor in monolayer MoS2nanosheet, *J. Phys. Condens. Matter*, 2017, **29**, 225501.

62  M.-K. Han, Y. Jin, D.-H. Lee and S.-J. Kim, Thermoelectric properties of Bi2Te3: CuI and the effect of its doping with Pb atoms, *Materials (Basel).*, 2017, **10**, 1235.

63  S. Ju, T. Shiga, L. Feng and J. Shiomi, Revisiting PbTe to identify how thermal conductivity is really limited, *Phys. Rev. B*, 2018, **97**, 184305.

64  A. Shafique and Y.-H. Shin, Thermoelectric and phonon transport properties of two-dimensional IV–VI compounds, *Sci. Rep.*, 2017, **7**, 1–10.

65  R. Yan, J. R. Simpson, S. Bertolazzi, J. Brivio, M. Watson, X. Wu, A. Kis, T. Luo, A. R. Hight Walker and H. G. Xing, Thermal conductivity of monolayer molybdenum disulfide obtained from temperature-dependent Raman spectroscopy, *ACS Nano*, 2014, **8**, 986–993.

66  Z. Jin, Q. Liao, H. Fang, Z. Liu, W. Liu, Z. Ding, T. Luo and N. Yang, A revisit to high thermoelectric performance of single-layer MoS 2, *Sci. Rep.*, 2015, **5**, 18342.

67  S. Kumar and U. Schwingenschlögl, Thermoelectric response of bulk and monolayer MoSe2and WSe2, *Chem. Mater.*, 2015, **27**, 1278–1284.